\documentclass[aps,prd,article]{revtex4}
\usepackage{amsmath}
\usepackage{amssymb}
\usepackage{amsthm}
\usepackage{array}
\usepackage{xy}
\bibliography{apsrev}
\begin{document}

\title{Trace Anomaly in Quantum Spacetime Manifold}
\author{Euro Spallucci}
\email{spallucci@trieste.infn.it} \affiliation{Dipartimento di
Fisica Teorica,
                  Universit\`a di Trieste
                  and INFN, Sezione di Trieste}
\author{Anais Smailagic}
\email{anais@ictp.trieste.it} \affiliation{ INFN, Sezione di
Trieste}
\author{Piero Nicolini}
\email{nicolini@cmfd.univ.trieste.it} \affiliation{Dipartimento di
Matematica e Informatica,
                  Universit\`a di Trieste
  and INFN, Sezione di Trieste}
\date{\today}

\begin{abstract}
In this paper we investigate the  trace anomaly in a spacetime
where single events are de-localized as a consequence of short
distance quantum coordinate fluctuations. We obtain a modified
form of heat kernel  asymptotic expansion
 which does not  suffer  from  short distance divergences.
Calculation of the trace anomaly is performed  using an IR
regulator in order to circumvent the absence of UV infinities   .\\
The explicit  form of the trace anomaly is presented and the
corresponding $2D$  Polyakov effective action  and   energy
momentum
tensor  are obtained. \\
The vacuum expectation value of the energy momentum tensor in the
Boulware, Hartle-Hawking and Unruh vacua is explicitly calculated
in a $rt$ section of a recently found, noncommutative inspired,
Schwarzschild-like solution of the Einstein equations.  The
standard short distance divergences in the vacuum expectation
values are regularized in agreement with the absence of UV
infinities removed by quantum coordinate fluctuations.
\end{abstract}

\maketitle
\section{Introduction}
Quantum gravity has been considered for a long time the would be
quantized version of General Relativity, or of any of its possible
extension. The success of String Theory has shown that the former
way is actually untenable and that a fully consistent unified
theory of all fundamental interactions, including gravity,
requires a deeper level of quantization. Rather than quantization
of fields propagating in a classical, non-dynamical, manifold it
is space-time itself that must be quantized. The boldest attempt
in this direction is offered by the still ``$M$-ysterious''
$M$-theory \cite{M}, where classical space-time coordinates are
expected to emerge as
classical eigenvalues of noncommuting quantum string coordinates. \\
An alternative  approach to space-time quantization is given by
Noncommutative Geometry where the quantum features are encoded
into a non-vanishing coordinate commutator. The technical
difficulty of dealing with coordinates which are not $c$-numbers
but ``operators'' is usually by-passed by introducing ordinary,
commuting, coordinates and shifting noncommutativity (NCY) in a
new multiplication rule, i.e. $\ast$-product, between functions.
This approach has been widely exploited and it can be summarized
in the following prescription: take commutative QFT results and
substitute ordinary function multiplication by $\ast$-product
multiplication. This prescription,
  however, leaves quadratic terms in the action unaffected, as the explicit
  form of the $\ast$-product leads, in this case, to surface terms only. Thus,
  free dynamics encoded into kinetic terms and Green functions remains the
same as in the commutative case. The presence of a non-commutative
product becomes relevant when more than two field variables couple
together. The $\ast$-product is a \textit{non local} operation
giving rise to non-planar contributions to Feynman diagrams at any
perturbative order.  These non-planar graphs introduces an
unexpected and unpleasant mixing of ultraviolet (UV) and infrared
(IR) divergences
  \cite{UV/IR}. So far there has
  been no solution to this problem and it has been accepted as an
unavoidable
  byproduct of NCY.  This is only one of the technical problems in a series
  which includes breaking of Lorentz invariance and problems with unitarity
  \cite{lorunitar}. \\
  Recently we have proposed an alternative approach to space-time
quantization
  based on the use of  coherent states of the quantum position operators
  $\textbf{x}^\mu$ \cite{ae1} satisfying

  \begin{equation}
   \left[\, \textbf{x}^\mu\ ,\textbf{x}^\nu\,
   \right]\ne 0\label{n0}
   \end{equation}

  The idea is that the non-vanishing commutator (\ref{n0})
  excludes a common basis in coordinate representation. The
  best one can do is to define mean values between appropriately chosen
states,
  i.e. coherent states. These mean values are the closest one can get to the
  classical commuting coordinates, since coherent states are minimum
uncertainty
  states.  In \cite{ae1} we applied this
  procedure by constructing ladder operators built of quantum
  coordinates only, in analogy to the usual phase space ladder operators
which
  are mixtures of coordinates and momenta. The use of mean values of quantum
  position operators as ``classical coordinates'' leads to the emergence
  of a \textit{quasi-classical} space-time manifold where the position of
any
  physical object is intrinsically uncertain. This uncertainty can be seen
either
  as a Gaussian cut-off in momentum space Green functions, or as a
substitution
  of position Dirac delta  with minimal width Gaussian function.\\
  It is important that the reader realizes that this cut-off is
  \textit{not put by hand} as an auxiliary regularization, but is
  direct consequence of the use of the position coherent states.
  In coherent states approach the problem
  of unitarity and Lorentz invariance can be handled properly and Gaussian
  cut-off renders the theory UV-finite. Furthermore no UV/IR mixing can
appear
  simply because there are no UV divergences.\\
  It should be interesting to investigate the modification of some relevant
  results of standard QFT within position coherent states formulation. \\
  One of the well known effects, in both gauge theories and gravity, is the
  anomalous quantum breaking of classical symmetries.
  Anomalies, i.e. non-vanishing divergences of quantum symmetry currents,
are
  usually extracted after suitable manipulations (~regularization~) of
divergent
  quantities. These divergences are produced by coincidence limit of various
  operator products and are consequences of the point like structure of
  commutative space-time. It is interesting to look at the anomaly problem
in
  NC space-time. Within the $\ast$-product approach the anomalies
  have been already investigated and, again, the only modification
  to standard results is the substitution of ordinary function
multiplication
  with $\ast$-multiplication \cite{staranomalies}.\\
  In the coherent states approach we do not expect the same result for two
  reasons: first, there are no UV divergences from which to extract
anomalies;
  second, there is no  $\ast$-product at all. Thus, either it could be
  that there are no anomalies
  at all, or there are modifications of the standard results different from
the
  one obtained in the $\ast$-product.\\
  The paper is organized as follows: in Sect.2 we show that the standard
trace
  anomaly can alternatively be extracted from IR divergences. This is a
  method followed in the case where there are  no UV divergent Feynman
diagrams,
  and thus suitable for our purposes. Secondly, we define
  heat-kernel and effective action for a gauge D' Alambertian operator in a
  quasi-classical space-time.
  At the end, we perform an explicit calculation of the  trace anomaly
  for massless fermions coupled to a background Abelian gauge field.
  Section 3 is devoted to the extension of the previous method to a curved
  space-time and to the consequent calculation of the gravitational trace
anomaly.
  As an example, we integrate the $2D$ trace anomaly and obtain the
  Polyakov effective action in a ``quantum'' string world manifold.
  In the Appendix we introduce a useful relation
  between a Gaussian function and a Dirac delta, used throughout the paper.

\section{Heat Kernel in flat, quasi-classical, space-time}

  Quantum space-time manifold is defined by the coordinate commutator

  \begin{equation}
   \left[\, \textbf{x}^\mu\ ,\textbf{x}^\nu\,
   \right]=i\,\Theta^{\mu\nu} \label{xcomm}
   \end{equation}

   where $\Theta^{\mu\nu}$ is a constant, antisymmetric, Lorentz tensor.
   In case of coordinates satisfying (\ref{xcomm}) the usual notion of
   classical
   space-time is lost and is replaced by a ``quantum'' geometry. Thus, we
face
   two levels of ``quantization'': one due to space-time uncertainty and the
   other due to matter. The interest for NCQFT comes from the interplay
between
   these two levels of quantization and the resulting physical effects.\\
   In even dimensional space-time $\Theta^{\mu\nu}$ can be brought to a
   block-diagonal form by a suitable Lorentz rotation leading to

   \begin{equation}
   \Theta^{\mu\nu} =\mathrm{diag}\left[\,
   \theta_1\,\mathbf{\epsilon}^{ab}\,
    \theta_2\,\mathbf{\epsilon}^{ab}\,
   \dots \theta_{d/2}\,\mathbf{\epsilon}^{ab}\,\right]
   \end{equation}

  with $\mathbf{\epsilon}^{ab} $ a $2\times 2$ antisymmetric Ricci
Levi-Civita
  tensor. It has been shown \cite{noi} that the resulting field theory is
  Lorentz invariant if all the $\theta$-parameters are coincident:
  $\theta_i\equiv \theta$.
  Indeed, Lorentz invariance requires space-time homogeneity and isotropy.
  The $\theta$ parameter is a UV cut-off in QFT on the quasi-classical
  space-time. At the same time,  $\theta$ has dimension $\left(\,
  \mathrm{mass}\,\right)^{-2}$, thus one would expect that its presence
  leads to an \textit{explicit breaking} of scale invariance.
  It will be shown that there is no explicit $\theta$-breaking since
  $\theta$ parameter always comes in a particular scale invariant
combination.
  Thus, scale anomaly is still a quantum mechanical
  effect in quasi-classical space-time manifold, but now endowed with
  $\theta$ corrections. \\
   We briefly list methods for calculating anomalies. First calculations
  dealt with axial current anomalous divergence which was obtained
  in perturbation theory
  by computing the appropriate Feynman diagram \cite{aj}, and later repeated
  in the case of trace of the energy momentum tensor \cite{duff}. Soon after

  Fujikawa noticed that anomalies in the path-integral formalism  are due
  to the non-invariance of the functional measure under classical symmetries
  \cite{fuji}.  Heat Kernel expansion \cite{dewitt} provides another very
  efficient method to extract divergent parts from quantum amplitudes
  \cite{iqg}. Alternatively, one can resort to $\zeta$-function \cite{zeta},
  or dimensional regularization \cite{dim}.
  As a result, all these different methods boil down to the introduction,
   at the quantum level, of some kind of regulating parameter, with
dimension
  of mass, thus breaking classical scale invariance of the model.\\
  In our case, the absence of UV divergences
  requires a suitable modification of the standard approach.
  Therefore, let us think of another reason to
  introduce  some  parameter with dimension of mass, with the
  hope that it would still lead to the appearance of an anomalous trace.
  Fortunately,  such a parameter
  is required to regulate IR divergences in massless QFTs. 
starting
  The  procedure
  is to define the path integral in an infrared regulated way and
functionally
  integrate away matter fields. The resulting functional determinant will
  depend upon the infrared regulating parameter. Finally, the regulated
trace
  of the vacuum averaged energy-momentum tensor is recovered after the
infrared
  cut-off is removed.  An eventual non-zero result will give us the desired
  trace anomaly.\\
   Let us apply this procedure in the case of
  a massless fermionic field $\psi\left(\, x\,\right)$ coupled to an
  Abelian background gauge field $A_\mu\left(\, x\,\right)$. The partition
  functional for the background field is given by

  \begin{equation}
  Z\left[\, A\,\right]= N\,
  \int D\bar\psi\,  D\psi \,\exp\left[\, -\int d^dx\, \bar\psi\, i\gamma^\mu
  D_\mu \, \psi\,\right]
  \end{equation}

  where $D_\mu\equiv \partial_\mu -i\,e \, A_\mu$ is the $U\left(\,1\
\right)$
  gauge covariant derivative.
  and the normalization constant $N$ is defined as

  \begin{equation}
  N^{-1}\equiv \int D\bar\psi\,  D\psi \,\exp\left[\, -\int d^dx\,
  \bar\psi\, i\gamma^\mu \partial_\mu \, \psi\,\right]
  \end{equation}
  With this choice $Z\left[\, 0\,\right]=1$.
  The effective action for the background field is obtained by integrating
  out the matter field $\psi$, which gives
  \begin{eqnarray}
  Z\left[\, A\,\right]&&=\frac{\mathrm{det}\left[\,
i\gamma^\mu\,D_\mu\,\right]}
  {\mathrm{det}\left[\, i\gamma^\mu\,\partial_\mu\,\right]}
  =\exp\left[\,
  \ln\, \mathrm{det}\left(\, i\gamma^\mu\,D_\mu\,\right)-
  \ln\,\mathrm{det}\left(\, i\gamma^\mu\,\partial_\mu\,\right)\,\right]
  \nonumber\\
  &&=
  \exp\left\{\, \frac{1}{2}\mathrm{Tr}\left[\,
  \ln\,\left(\,i\gamma^\mu\,D_\mu\,\right)^2  -
  \ln\,\left(\,-\Box\,\right)\,\right]\,\right\}\label{uno}
\end{eqnarray}

  As the starting theory is massless, let us take care of infrared
  divergences by introducing a fictitious mass parameter, $\mu$.
  This leads to the \textit{ IR regularized effective action}:

\begin{equation}
  \Gamma\left[\, A\,\right]_{IR-reg}\equiv
  \int d^d x \, L\left[\, A\,\right]=\frac{1}{2}\,\lim_{\mu^2\to 0}
  \mathrm{Tr}\left[\, \ln\,\left(\,\Delta +\mu^2\,\right)
  - \ln\,\left(\,-\Box\,\right)\,\right]\label{eff}
  \end{equation}

  where, $\Delta\equiv \left(\,i\gamma^\mu\,D_\mu\,\right)^2$.
  We are going to show that IR regularization procedure can be used to
obtain
  the trace anomaly.  Being  an
  auxiliary parameter, and not a physical mass, $\mu$ will be set to zero at
  the end of calculations. It turns out that finite terms survive
  leading to the \textit{Trace Anomaly}. \\
  To prove this statement, let us consider an infinitesimal rescaling of the
  Euclidean metric:
  $\delta_{\mu\nu}\longrightarrow \left(\,1 -2\epsilon\,\right)
\delta_{\mu\nu}$.
  Both D' Alambertian operators in (\ref{eff}) scale with weight $2$
  and $L$ varies as

  \begin{equation}
  dL= \frac{\partial L}{\partial \epsilon} \, d\epsilon=
  2\mu^2\, \mathrm{Tr}\left[\, \frac{1}{1+2\epsilon}\,\,
  \frac{1}{\mu^2 -\left(\,1+2\epsilon\,\right) \Delta}\,\right]\, d\epsilon
  \label{riscalo}
  \end{equation}

  The trace anomaly represents the response of the effective Lagrangian
under
  rescaling, namely

  \begin{equation}
\langle \, 0\, \vert\, T^\nu{}_\nu\, \vert\, 0\, \rangle =
\lim_{\epsilon\to 0}\left(\, \frac{\partial L}{\partial\epsilon
}\,\right)\label{tadef}
\end{equation}

  By taking into account (\ref{riscalo}) we get

  \begin{eqnarray}
  \lim_{\epsilon\to 0}\frac{\partial L}{\partial\epsilon }&&=
2\mu^2\,\mathrm{Tr}
  \,\left[\,\frac{1}{ - \Delta^2 +\mu^2}\,\right]_{\mu^2=0}\nonumber\\
  &&=2\lim_{\mu^2\to 0} \mu^2\, \left(\,
  \frac{\partial L}{\partial \mu^2 }\,
\right)
  \end{eqnarray}

  Thus, we can express the trace anomaly as

  \begin{equation}
\langle \, 0\, \vert\, T^\nu{}_\nu\, \vert\, 0\, \rangle =
\left(\, \mu^2\, \frac{\partial L}{\partial
\mu^2}\,\right)_{\mu^2=0}\label{ta}
\end{equation}

Equation (\ref{ta}) is the definition of the trace anomaly in
terms of the IR regulator. It shows that only the $\mu$-dependent
part of  $L$ can contribute to the anomaly. Thus, in what follows
we shall drop the field independent (~infinite~) term $\ln
\left(\, -\Box\,\right)$.\\
Our goal is to investigate the effects of quantum space-time
manifold on the effective action and trace anomaly. Recently two
of us have proposed a way of introducing space-time fluctuations
in QFT formalism using coordinate coherent states \cite{ae1}. The
motivations are discussed in the introduction. The outcome of our
approach is that the uncertainty in \textit{coordinate}
representation is seen through Gaussian
function of width $\sqrt{2\theta}$ in place of Dirac delta function.\\
In view of the above prescription,  Green function equation for
the square of the Dirac operator becomes

\begin{equation}
\left(\,\Delta +\mu^2\,\right)\, G_\theta\left(\, x\ ,y\,\right)=
\rho_\theta\left(\, x\ ,y\,\right)\ ,\label{green}
\end{equation}

where $\rho_\theta\left(\, x\ ,y\,\right)$ is the above mentioned
Gaussian source:

\begin{equation}
\rho_\theta\left(\, x\ ,y\,\right)= \frac{1}{\left(\,
4\pi\,\theta\,\right)^{d/2} } \exp\left( -\frac{\left(\, x -
y\,\right)^2}{4\theta}\,\right)
\end{equation}

Although there are no \textit{physical} point like sources, one
can use delta-functions in the intermediate steps thanks to a
useful representation of  Gaussian of width $\theta$ in terms of
the Dirac delta as (~see Appendix A~):

\begin{equation}
e^{\theta\,\Box_x}\, \delta^{(d)}\left(\, x-y\,\right)
=\rho_\theta\,\left(\, x\ ,y\,\right)\label{dirac}
\end{equation}

The solution of Eq.(\ref{green}) is given by

\begin{equation}
G_\theta\left(\, x- y\,\right)= \int d^d z \,G_0\left(\,
x-z\,\right) \rho_\theta\left(\, z-y\,\right)\ ,
\end{equation}

where $G_0\left(\, x-z\,\right)$ is the commutative Green function
satisfying

\begin{equation}
\left(\, \Delta + \mu^2\,\right)\, G_0\left(\, x-z\,\right)=
\delta^{(d)}\left(\, x-z\,\right)
\end{equation}

The commutative Green function $G_0\left(\, x-z\,\right)$ can be
expressed through the commutative Heat Kernel $K_0\left(\, x-z\ ;
s\,\right)$ as

\begin{equation}
G_0\left(\, x-z\,\right)=\int_0^\infty ds\, K_0\left(\, x-z\ ;
s\,\right)
\end{equation}

Thus, we exploit the relation between modified and commutative
quantities to write

\begin{eqnarray}
G_\theta\left(\, x- y\,\right) &&=
  \int_0^\infty ds\,\int d^d z \, K_0\left(\, x-z\ ; s\,\right)
\rho_\theta\left(\, z-y\,\right)\nonumber\\
&&\equiv \int_0^\infty ds\,K_\theta\left(\, x\ ,y\ ; s\,\right)
\end{eqnarray}

where we arrived to the definition of the \textit{modified Heat
Kernel } which can be further re-written as

\begin{eqnarray}
K_\theta\left(\, x\ ,y\ ; s\,\right) &&=\int d^d z \, K_0\left(\,
x-z\ ; s\,\right)\, e^{\theta\, \Box_z}\,\delta^{(d)}\left(\,
x-z\,\right)
\nonumber\\
&&=e^{\theta\,\Box_x}\, K_0\left(\, x\ ,y\ ; s\,\right)\label{def}
\end{eqnarray}

in obtaining (\ref{def}) we have exploited well known integration
properties
of the Dirac delta.\\
It can be shown that the noncommutative kernel (\ref{def})
satisfies the standard heat equation but with modified initial
condition which takes into account the fuzziness of the space-time

\begin{eqnarray}
&&\Delta\, K_\theta\left(\, x\ ,y\ ; s\,\right)=
\frac{\partial}{\partial s}\,
K_\theta\left(\, x\ ,y\ ; s\,\right)\nonumber\\
&& K_\theta\left(\, x\ ,y\ ; 0\,\right)=\rho_\theta\left(\, x\
,y\,\right) \label{heat}
\end{eqnarray}

\subsection{Asymptotic expansion}

In order to calculate the trace anomaly in quantum space-time one
starts from Eq. (\ref{def}). Firstly, we exploit the asymptotic
expansion of the kernel in terms of the Seeley-DeWitt coefficients
\footnote{For  various problems and applications  of the Heat
Kernel expansion see \cite{hk}.}

\begin{equation}
K_\theta\left(\, x\ ,y\ ; s\,\right)= e^{\theta\,\Box_x}\,\left[\,
\frac{1}{\left(\, 4\pi s\,\right)^{d/2}} e^{-\mu^2\, s-\left(\, x
- y \right)^2/4s} \, \sum_{n=0}^\infty\, s^n\, a_n\left(\, x\
,y\,\right)\,\right]
  \label{expansion}
\end{equation}

Using the property that the first coefficient is the trace of the
identity matrix we can write (\ref{expansion}) as

\begin{eqnarray}
K_\theta\left(\, x\ ,y\ ; s\,\right)=&& a_0\, e^{\theta\,\Box_x}\,
\frac{1}{\left(\, 4\pi s\,\right)^{d/2}}\,
e^{-\mu^2\, s-\left(\, x - y\,\right)^2/4s}\nonumber\\
  +&&
\frac{1}{\left(\, 4\pi s\,\right)^{d/2}}\,
e^{\theta\,\Box_x}\,\left[\, e^{-\mu^2\, s-\left(\, x -
y\,\right)^2/4s} \, \sum_{n=1}^\infty\, s^n\, a_n\left(\, x\
,y\,\right)\,\right]
\end{eqnarray}

The second term can be manipulated using Fourier transform of the
coefficient as follows

\begin{eqnarray}
&& e^{\theta\,\Box_x}\,\int \frac{d^d p}{(2\pi)^d}\, e^{-s p^2 +
ip \left(\, x - y\,\right)}
\, \sum_{n=1}^\infty\, s^n\, a_n\left(\, x\ ,y\,\right)=\nonumber\\
&& \int \frac{d^d p}{(2\pi)^d}\, \int \frac{d^d q}{(2\pi)^d}\,
e^{-s \, p^2 -\theta\, \left(\, p+q\,\right)^2 +
i\left(\,p+q\,\right)\, x -i\, p\, y}
\, \sum_{n=1}^\infty\, s^n\,  a_n\left(\, q\ ,y\,\right)=\nonumber\\
&&= \frac{e^{-\left(\, x-y\,\right)^2/4\left(\,s + \theta\,
\right)}}{\left[\, \left(\, 4\pi\,\right)\,\left(\,
s+\theta\,\right)\,\right]^{d/2} }\, \, \int \frac{d^d
q}{(2\pi)^d}\,
  e^{-q^2\,\theta\, s/\left(\, \theta +s\,\right)}\,
  e^{i\, q\, \left[\, x + x\,\theta/ \left(\, s+\theta\,\right)
  - y\,\theta/ \left(\, s+\theta\,\right)\,\right]}\,
  \sum_{n=1}^\infty\, s^n\,  a_n\left(\, q\ ,y\,\right)\nonumber
  \\
&&
\end{eqnarray}

Thus, we arrive at the following asymptotic expansion for the Heat
Kernel

\begin{equation}
K_\theta\left(\, x\ ,y\ ; s\,\right)=\rho_{\theta+s}\left(\,
x-y\,\right)\, e^{-\mu^2\,s}\, \left[\, a_0 + \int \frac{d^d
q}{(2\pi)^d}\,
  e^{-q^2\,\theta\, s/\left(\, \theta +s\,\right)}\,
  e^{i\, q\, \left[\, x + \left(\,x-y\,\right)\,\theta\, s /
  \left(\, s+\theta\,\right)\,\right]}
  \, \sum_{n=1}^\infty\, s^n\,  a_n\left(\, q\ ,y\,\right)\,\right].
\end{equation}

In most of the actual applications of the Heat Kernel method one
needs the trace of the kernel, i.e. $   K_\theta\left(\, x\ , x\ ;
s\,\right)$:

\begin{eqnarray}
K_\theta\left(\, x\ , x\ ; s\,\right) &&=\rho_{\theta+s}\left(\,
0\,\right)\,e^{-\mu^2\, s}\, \left[\, a_0 + e^{\left[\, \theta\,
s/\left(\, \theta +s\,\right)\,\right]\, \Box}\, \int \frac{d^d
q}{(2\pi)^d}\,
  e^{i\, q\,  x }\, \sum_{n=1}^\infty\, s^n\,  a_n\left(\, q\
,x\,\right)\,\right]
\nonumber\\
&&=\rho_{\theta+s}\left(\, 0\,\right)\,e^{-\mu^2\, s}\,\left[\,
a_0 + e^{\,\left[\,\theta\, s/\left(\, \theta
+s\,\right)\,\right]\, \Box}\, \sum_{n=1}^\infty\, s^n\,
a_n\left(\, x\ ,x\,\right)\,\right]
\end{eqnarray}

We find the modified version of the asymptotic expansion for the
Heat Kernel to be

\begin{equation}
K_\theta\left(\, x\ , x\ ; s\,\right)= \frac{1}{\left[\,\left(\,
4\pi\,\right)\,\left(\, s+\theta\,\right)\,\right]^{d /2} }
\,e^{-\mu^2\, s}\,\left[\, a_0 + e^{\,\left[\, \theta\, s/\left(\,
\theta +s\,\right)\,\right]\, \Box}\, \sum_{n=1}^\infty\, s^n\,
a_n\left(\, x\ ,x\,\right)\,\right] \label{trace}
\end{equation}

\subsection{Heat Kernel representation of the Effective Action in
quasi-classical space-time}

To obtain the form of the effective action in terms of the Heat
Kernel, one \textit{formally} varies the differential operator
$\Delta$ and computes the corresponding variation of the effective
action  as

\begin{eqnarray}
\delta\,\Gamma && = \frac{1}{2}\, \delta\,\int d^dx\,
\mathrm{Tr}\ln\,\left(\,\Delta+\mu^2\,\right)
\nonumber\\
&&=\frac{1}{2}\,\int
d^dx\,\mathrm{Tr}\left[\,\frac{1}{\Delta+\mu^2}\,
\rho_\theta\left(\, x\ , x\,\right)\,\right]\, \delta\, \Delta
\nonumber\\
&&=\frac{1}{2}\,\int d^dx\,\mathrm{Tr}\int_0^\infty ds\,
  e^{-s\,\left(\,\Delta+\mu^2\,\right) }\, e^{\theta\, \partial_x^2}\,
  \delta^{(d)}\left(\, x\ ,x\,\right)  \, \delta\, \Delta
\end{eqnarray}

where we have taken into account (\ref{dirac}) and momentarily
dropped the limit $\mu^2\to 0$. The operator $\Delta$ can be
always cast in the form

\begin{equation}
\Delta= -D^2 + X\left(\, x \,\right)
\end{equation}

where, $D^2$ is the covariant D' Alambertian with respect to the
local symmetry that one wants to maintain at the quantum level.
The exact form of the kernel of $\Delta$ is not known, but it is
sufficient to consider only  the asymptotic form (\ref{trace}).
  Asymptotic expansion (\ref{trace}) is valid for the values of $s$ for
which
  the fields contained in $X\left(\, x \,\right)$ are slowly varying, so
that
  the condition

  \begin{equation}
\Box\, X\left(\, x \,\right)=0 \longrightarrow \left[\, \Box\ ,
X\left(\, x \,\right)\,\right]=0
\end{equation}

  is satisfied. This assumption enables  to
``integrate''  the functional variation of the effective action:

\begin{eqnarray}
\delta\,\Gamma && =\frac{1}{2}\, \int d^dx\,
\mathrm{Tr}\int_0^\infty ds\,
  e^{-\left(\, s+\theta\,\right)\,\Box -\left(\, s+\theta\,\right)
  \left(\, X(x) +\mu^2\,\right) + \theta\,\left(\, X(x) +\mu^2 \,\right) }\,
  \,\delta^{(d)}\left(\, x\ ,x\,\right)  \, \delta\, \Delta\nonumber\\
&& =\frac{1}{2}\, \int d^dx\, \mathrm{Tr}\int_0^\infty ds\,
  e^{-\left(\, s+\theta\,\right)\,\left(\, \Delta +\mu^2\,\right)
  + \theta\,\left(\, X(x) +\mu^2 \,\right) }\,
  \,\delta^{(d)}\left(\, x\ ,x\,\right)  \, \delta\, \Delta\nonumber\\
&& =-\frac{1}{2}\, \int d^dx\,
\delta\left[\,\mathrm{Tr}\int_0^\infty \frac{ds}{s+\theta}\,
  e^{\theta\,\Box}\,e^{-s\,\left(\, \Delta +\mu^2\,\right)}\,
  \,\delta^{(d)}\left(\, x\ ,x\,\right)\,\right]
\label{vargamma}
\end{eqnarray}

The above manipulation enable us to write the final form of the
modified effective action as

\begin{equation}
\Gamma= -\frac{1}{2}\lim_{\mu^2\to 0}\,\int d^dx\, \int_0^\infty
ds\,\frac{\, e^{-s\,\mu^2}}{s+\theta}\,e^{\theta\,\Box}\,
K_0\left(\, x\ , x\ ; s\,\right)\label{nceff}
\end{equation}

The effective action (\ref{nceff}) is  ultraviolet
\textit{finite}. Short distance divergences have been removed in
the quantum space-time. This is what Gaussian functions,
representing position uncertainty, do to ordinary QFT \cite{ae1},
\cite{noi} .

\subsection{Trace Anomaly calculation}

Usually anomaly are calculated from the divergent part of the
asymptotic expansion of the effective action. However, in
quasi-classical space-time these divergences have been cured and
one cannot follow the same path. In Section(2) we have shown that,
alternatively, anomaly can be calculated through the IR regulator
(\ref{ta}). Thus, let us calculate the derivative of the effective
Lagrangian
  (\ref{nceff}) with respect to $\mu^2$:

\begin{eqnarray}
\mu^2\, \frac{\partial L^{eff.}}{\partial \mu^2} &&=\frac{1}{2}
\left(\,\frac{\mu^2}{4\pi}\,\right)^{d/2}\,\times\nonumber\\
&&\int_0^\infty d\tau\, \frac{\tau\, e^{-\tau}} {\left[\, \left(\,
\tau+\mu^2\, \theta\,\right)\,\right]^{1+ d/2}}\,
\mathrm{Tr}\left[\, a_0 + e^{\,\left[\, \theta\,
\tau/\left(\,\mu^2 \theta +\tau\,\right)\,\right]\,\Box}
  \,\sum_{n=1}^{d/2}\,\left(\,\frac{\tau}{\mu^2}\,\right)^n\,
  a_n\left(\, x\,\right)\,\right]\label{ir}
\end{eqnarray}

In the limit $\mu^2\to 0$ the only non-vanishing and finite
contribution comes from the $n=d/2$ term in the sum. The IR
divergent terms of (\ref{ir}) are subtracted in the usual manner
and do not contribute to the anomaly. Therefore, the trace anomaly
turns  out to be

\begin{eqnarray}
\langle\, 0\, \vert\, T^\nu{}_\nu\,\vert\, 0\,\rangle &&\equiv
2\lim_{\mu^2\to 0}\left(\,
  \mu^2\, \frac{\partial L^{eff}}{\partial \mu^2}\,\right)\nonumber\\
&&= \lim_{\mu^2\to 0}\frac{\left(\,\mu^2\,\right)^{d/2}}{\left(\,
4\pi\,\right)^{d/2}}\, \mathrm{Tr}\int_0^\infty d\tau\,
\frac{\tau\, e^{-\tau} }{ \tau^{1+ d/2}
}\,\left(\,\frac{\tau}{\mu^2}\,\right)^ {d/2}
\,e^{\,\theta\,\Box}\, a_{d/2}\left(\, x\,\right)\nonumber\\
&&=\frac{1}{\left(\, 4\pi\,\right)^{d/2}}\, e^{\,\theta\,\Box_x}\,
  \mathrm{Tr}\, a_{d/2}\left(\, x\,\right)\label{atrace}
\end{eqnarray}

The result (\ref{atrace}) holds in any dimension and is the
modified version of the usual trace anomaly.
  As an explicit example, let us consider $D$ trace anomaly of massless
fermionic
  matter in an Abelian gauge field background. It is found to be

\begin{equation}
\langle\, 0\, \vert\, T^\lambda_{\lambda}\,\vert\, 0\,\rangle =
\frac{e^2\, N_F}{12\,\left(\, 4\pi\,\right)^2}\,
e^{\,\theta\,\Box}\,
  F_{\mu\nu}\, F^{\mu\nu} \label{noto}
\end{equation}

where, $N_F$ is the total number of fermionic degrees of freedom.
Equation (\ref{noto}) displays in a clear way how space-time
quantum fluctuations, described using coordinate coherent states,
affects the standard result through the new term
$e^{\,\theta\,\Box}$.
%

\section{Trace Anomaly in curved space-time}
So far, we have considered the trace anomaly in quasi-classical
flat space-time. In view of the role of conformal invariance, and
its breaking, in modern cosmological problems \cite{eig}, we would
like to extend
the above consideration to the curved quasi-classical space-time.\\
It is important to understand what modification will be sufficient
in the passage from classical to quasi-classical geometry. We have
already explained the effect of fluctuating quantum coordinates on
the Dirac delta function and this applies equally in curved
space-time. Furthermore, the question of covariance has to be
properly incorporated in operators connecting Gaussian and Dirac
delta-functions. Thus, the curved modified space-time version of
(\ref{green}) is

\begin{equation}
\left(\,\Delta +\mu^2\,\right)\, G_\theta\left(\, x\ ,y\,\right)=
e^{\theta\, \nabla_\mu\, \nabla^\mu } \frac{1}{\sqrt g}\,
\delta^{(d)}\left(\, x\ ,y\,\right)\ ,\label{cgreen}
\end{equation}

where $\nabla_\mu$ is the  generally covariant derivative in the
metric $g_{\mu\nu}$.We would like to point out that the role
  of the IR regulator  in (\ref{cgreen}) with respect to global scale
  transformations  has been explained previously, and it remains such
  in case of local Weyl symmetry. We are only  interested in quantum
  (~anomalous~) Weyl symmetry breaking that will, eventually,
  remain after the limit $\mu^2\to 0$ is taken. On the other hand, the
covariant
  D' Alambertian, is not  Weyl covariant in arbitrary dimension.
  We  modify the D'Alambertian  to achieve Weyl covariant form as

\begin{equation}
\nabla_\mu\, \nabla^\mu\longrightarrow  \nabla_\mu\, \nabla^\mu
-\xi_d \, R \equiv \nabla^2_w\ , \qquad \xi_d\equiv
\frac{1}{4}\frac{d-2}{d-1}
\end{equation}

which transforms  under $g_{\mu\nu}=\omega^2\, \hat{g}_{\mu\nu}$
as $\hat \nabla^2_w =\omega^{-2}(x)\, \nabla^2_w $.
  $\theta$ as a component of the matrix $\Theta^{\mu\nu}$
  is a constant with dimension of length squared. 
For this reason the  exponent in (\ref{cgreen}) is not
Weyl invariant  unless one expresses $\theta$ in terms of
$\Theta^{\mu\nu}$ as

\begin{equation}
\theta\longrightarrow \theta\left(\, x\,\right)\equiv
\sqrt{\frac{1}{d}\, g_{\mu\alpha}(x)\, g_{\nu\beta}(x)\,
\Theta^{\mu\nu}\, \Theta^{\alpha\beta} }
\end{equation}

and such $\theta\left(\, x\,\right)$ transforms under Weyl
rescaling as $\theta\left(\, x\,\right)=  \omega^{2}(x)  \hat
\theta\left(\, x\,\right)$. Now, we can safely claim the Weyl
invariance of the exponent on the r.h.s. of (\ref{green}). 
To avoid any confusion, we remark that $\Theta^{\mu\nu}$ remains
a constant tensor while spacetime dependence enters  in the
\textit{invariant scalar} $\theta\left(\,x\,\right)$ only through the
metric tensor. As a consequence there will be additional contributions
to the energy-momentum tensor due to the variation of 
$\theta\left(\,x\,\right)$.\\
We calculate the variation of the curved space-time extension of the
effective action (\ref{vargamma}) as

\begin{eqnarray}
\delta\,\Gamma
  && =\frac{1}{2}\, \int d^dx\sqrt{g}\, \mathrm{Tr}\int_0^\infty ds\,
  e^{-\left(\, s+\theta\left(\, x\,\right)\,\right)\,\nabla^2
  -s\left[\,\left(\, \beta -\xi_d\,\right)\,R + X(x) +\mu^2\,\right]}\,
  \frac{1}{\sqrt{g}}\,\delta^{(d)}\left(\, x\ ,x\,\right)  \, \delta\,\Delta
  \nonumber\\
&& =-\frac{1}{2}\,\delta \int d^dx\sqrt{g}\,
\left[\,\mathrm{Tr}\int_0^\infty \frac{ds}{s+\theta\left(\,
x\,\right)}\, e^{\theta\left(\, x\,\right)\, \nabla^2_w }
\,e^{-s\,\left(\, \Delta +\mu^2\,\right)}\,
  \frac{1}{\sqrt{g}}\,\delta^{(d)}\left(\, x\ ,x\,\right)\,\right]
\label{ccvar}
\end{eqnarray}

In this manner, one obtains the generally covariant form of the
heat kernel :

\begin{eqnarray}
K_\theta\left(\, x\ , x\ ; s\,\right)&&\equiv e^{\theta\left(\,
x\,\right)\,\, \nabla^2_w }\, K\left(\, x\ , x\ ; s\,\right)
\nonumber\\
&&=\frac{ e^{-\mu^2\, s }}{\left[\,\left(\, 4\pi\,\right)\,
\left(\, s+\theta\left(\, x\,\right)\,\right)\,\right]^{d/2} }\,
\left[\, a_0\,e^{ \frac{s\theta\left(\, x\,\right)}{ s+s\theta
\left(\, x\,\right) }   \,\xi_d\, R}
    + e^{\,   \frac{s\theta\left(\, x\,\right)}{ s+s\theta\left(\, x\
\right) }
   \, \nabla^2_w}\, \sum_{n=1}^\infty\, s^n\,  a_n\left(\,
x\ ,x\,\right)\,\right]\nonumber\\
&&
\end{eqnarray}

Following the same steps as in (\ref{atrace}) we find the trace
anomaly to be
\begin{eqnarray}
\mu^2 \,\frac{\partial L}{\partial\mu^2 }&&=\mu^2 \,\mathrm{Tr}\,
G\left(\ x\ ,
x\,\right)\nonumber\\
&&=\mu^2 \,\mathrm{Tr}\,\left[\, \frac{1}{\Delta +\mu^2}\,
e^{\frac{s\theta\left(\, x\,\right)}{ s+s\theta\left(\, x\,\right)
} \,\nabla^2_w }\, \frac{1}{\sqrt g}\, \delta^{(d)}\left(\, x\
,x\,\right)  \,\right]
\end{eqnarray}

\begin{equation}
\langle\, 0\, \vert\, T^\nu{}_{\nu}\,\vert\, 0\,\rangle =
\frac{1}{\left(\, 4\pi\,\right)^{d/2}}\, e^{\theta\left(\,
x\,\right)\,\nabla^2_w }\,
  \mathrm{Tr}\,\left[\, a_{d/2}\left(\, x\,\right)\,\right]\label{ctrace}
\end{equation}

\subsection{Anomalous Effective Action}

Although the heat kernel method formally allows to calculate
effective action it is possible in practice only if the complete
Kernel is known. Since, this is never the case in QFT, one limits
himself only to the calculation of the first few terms in the
asymptotic expansion. The alternative is the calculation of few
Feynman diagrams and covariantization of the result, but it turns
out to be tedious and lengthy. The quickest way is to integrate
the anomaly equation which in commutative two dimensions leads to
the Polyakov action. We find it the most convenient to follow the
last approach since the modification due to noncommutativity can
be integrated without big hassle to find

\begin{equation}
S= \frac{D-26}{48\pi}\int d^2x \sqrt{g}\, R\, \frac{1}{\Box}\,
  e^{ \theta\left(\, x\,\right)\,\Box}\, R \label{polya}
\end{equation}

It turns out that the effective action (\ref{polya}) generalizes
well known $2d$ anomalous effective action in a relatively simple
way. The reason is that the exponent was introduced in a Weyl
invariant way to avoid any explicit classical symmetry breaking.
This condition has nothing to do with the effective action but
turns out to be crucial in order to be able to integrate the
anomaly equation (\ref{ctrace}). To be more precise the trace of
vacuum energy momentum tensor couples to the Weyl degree of
freedom. The anomaly comes from the term in the effective action
which is linear in the Weyl degree of freedom which is already
contained in the Ricci scalar curvature $R$. No other Weyl
dependence should be present. In this case this is true due to the
construction of the exponent.  Thus, the anomaly (\ref{ctrace}) is
a purely quantum mechanical
effect.\\
  Action (\ref{polya}) can be cast in the local form
\begin{equation}
S= \int d^2x \sqrt{g}\, \left[\alpha\, \phi \, e^{
\frac{\theta}{2}\,\Box } R -\frac{1}{2} \nabla_\mu \phi
\,\nabla^\mu \phi\, \right]
  \label{local}
\end{equation}

  where the auxiliary field $\phi$ must solve the field equation

  \begin{equation}
  \Box\phi=-\alpha\, e^{ \frac{\theta}{2} \, \Box} \,R
  \label{aux}
  \end{equation}

  and
  $\alpha\equiv\sqrt{(26-D)/24\pi}$.\\
  In $2D$ the above formula simplifies due to $\theta^{\mu\nu}=
\theta\, \varepsilon^{\mu\nu}$  where $\varepsilon^{\mu\nu}$ is
the totally anti-symmetric symbol such that $g_{\mu\alpha}
g_{\nu\beta}\, \varepsilon^{\mu\nu}\, \varepsilon^{\alpha\beta}=
\mathrm{Det}(\, g_{\rho\tau}\,)$.\\
  The energy momentum tensor following from the
  Polyakov effective active is to be understood as
  the vacuum expectation value of  a quantum energy
  momentum tensor since it already incorporates
  anomaly induced  quantum contributions.  It is given by

\begin{eqnarray}
  T_{\mu\nu} &&= \nabla_\mu \,\phi \nabla_\nu\,\phi -
\frac{1}{2}g_{\mu\nu}\left(\nabla_\rho
\phi\right)^2+\nonumber\\
  &&-\alpha\,\theta\, \phi \left(
\nabla_\mu\, \nabla_ \nu  -\frac{1}{2}\, g_{\mu\nu}\,\square
\right)\,
  e^{\frac{\theta }{2}\sqrt{-g}\square }\, R-2\alpha \left(
g_{\mu \nu }\square -\nabla _{\mu }\nabla _{\nu }\right)
e^{\frac{\theta}{2} \sqrt{-g}\Box }\phi \label{vac}
  \end{eqnarray}

In order to perform an explicit calculation of the vacuum
energy-momentum tensor in a $2D$ black hole metric, the standard
precedure amounts to consider as a background geometry the 
$rt$ section of the corresponding  $4D$ black hole metric.
This is known as the ``s-wave approximation'' \cite{swave}.
The use of Polyakov action, derived from the anomaly, already
incorporates quantum effects.  Thus, it provides a description
of $2D$ quantum black holes \cite{balbi2} and hopefully contains
useful information about black hole evaporation in $4D$. Motivated by
this argument we want to investigate coordinate non-commutativity
in this framework. In order to do so, we adopt the $rt$ section of a 
recently calculated, spherically symmetric  solution of the non-commutative 
inspired Einstein equations in $4D$ \cite{last}.   In this way we keep
track of the effects of noncommutativity both on the metric itself
and on the matter fields propagating in this
background geometry.\\
The  $rt$ section of the  metric obtained in \cite{last}, is

\begin{eqnarray}
&& ds^2=-f\left(r\right)\, dt^{2}+\frac{1}{f\left(r\right)}\, dr^{2}\\
&& f\left(r\right)= 1-\frac{4M}{r\sqrt{\pi }}\gamma \left(
\frac{3}{2}, \frac{r^{2}}{4\theta }\right) \equiv
1-\frac{2M\left(\, r\,\right)}{r}\label{nostra} \\
&&\sqrt{-g}=1\\
&& \gamma\left(\, \frac{3}{2}\ ,\frac{r^2}{4\theta }\, \right)
\equiv \int_0^{r^2/4\theta} dt t^{1/2} e^{-t}
\end{eqnarray}

where, $\gamma\left( \frac{3}{2},\frac{r^{2}}{4\theta }\right)$ is
a upper incomplete Gamma function whose main properties are listed
in Appendix B. The  $2D$ Ricci scalar   turns out to be

\begin{equation}
R=-f^{\prime \prime }=\frac{8M}{\sqrt\pi\, r^3}\left[\,
\gamma\left( \frac{3}{2},\frac{r^{2}}{4\theta }\right)-
\frac{r^5}{16 \theta^{5/2}}\,  e^{-r^2/4\theta}\,\right]
\end{equation}

where  ``prime''  stands for $d/dr$ . \\
The solution for the auxiliary field given by (\ref{aux}) is

\begin{equation}
\phi =e^{\frac{\theta }{2}\square }\left(\,  A\, t+\alpha\, \ln
f+B\, r^{\ast}+C\, \right)\equiv e^{\frac{\theta}{2}\square
}\phi^0 \label{fi}
\end{equation}

where $A,$ $B$ and $C$ are constants, while $r^{\ast }$ is the
tortoise coordinate defined as

\begin{equation}
r^\ast=\int \frac{dr}{f(r)}.
\end{equation}

 Fixing integration constants in (\ref{fi}) leads to
the choice of different  vacua  \cite{balbi} .\\
 The components of (\ref{vac}) are

\begin{eqnarray}
T_{tt} &=&\left(\partial_t\phi\right)^2 +
\frac{f}{2}\left(\partial_t\phi\right)\nonumber\\
&-&\frac{\alpha \theta }{2}\phi f\left( \,f^{\prime
}\partial_{r}+\square \right) e^{\frac{\theta }{2}\square
}R+2\alpha \left( f\square +\nabla_{t}\partial _{t}\right)
e^{\frac{\theta }{2}\square
}\phi  \\
T_{rr} &=& \left(\partial_r\phi\right)^2 -
\frac{1}{2f}\left(\partial_r\phi\right)\nonumber\\
&+& \frac{\alpha \theta }{2}\phi \left( -2\nabla_r\partial_r
+\frac{1}{f}\square \right) e^{ \frac{\theta }{2}\square
}R-2\alpha \left( \frac{1}{f}\square -\nabla
_{r}\partial_{r}\right) e^{\frac{\theta }{2}\square }\phi \\
T_{tr} &=&\left(\partial_t\phi\right) \left(\partial_r\phi\right)
-\alpha \theta\, \phi \nabla_r\partial_t  e^{\frac{\theta
}{2}\square }R+        2\alpha
  \nabla_{r}\partial _{t} e^{\frac{\theta }{2}\square}\phi
\label{comp}
\end{eqnarray}

where $\square =f^{\prime }\partial _{r}+f\partial
_{r}^{2}-f^{-1}\,\partial_t^2$.

  \subsection{Choice of vacua in quasi-classical metric}

The components of the energy-momentum tensor (\ref{comp}) can be
conveniently written as

\begin{eqnarray}
 T_{tt}  &=& \frac{{\dot \phi ^2 }}{2} + \frac{{f^2 }}{2}\phi '^2  + 2\alpha
f^2\left( {\tilde \phi '' + \frac{{f'}}{{2f}}\tilde \phi '}
\right) -
\frac{{\alpha \theta }}{2}\phi f^2 \tilde R'' \nonumber\\
 T_{rr}  &=& \frac{{\dot \phi ^2 }}{{2f^2 }} + \frac{{\phi '^2 }}{2} -
\alpha \frac{{f'}}{f}\tilde \phi ' - \frac{{\alpha \theta
}}{2}\phi \tilde R'
\\
 T_{tr}  &=& \dot \phi \left( {\phi ' - \alpha \frac{{f'}}{f}} \right)
\nonumber
\end{eqnarray}

 We  introduced  field redefinitions
$\tilde\phi=e^{\theta\square}\phi^0$ and $\tilde R=
e^{\theta\square/2}R$ in order to have covariant $\square$ act
only on scalars.The light cone components of momentum-energy
tensor are useful for choice of  different vacua and are given by

 \begin{eqnarray}
4T_{uu} \equiv  T_{tt}  + f^2 T_{rr}  - 2fT_{tr}&=&   \dot \phi ^2
+ f^2 \phi ^2  + 2\alpha f^2 \tilde \phi '' - \alpha \theta \phi
f^2 \tilde R'' - 2\dot
\phi \left( {f\phi ' - \alpha f'} \right)\nonumber \\
 &=& \left( {\dot \phi  - f\phi '} \right)^2  + 2\alpha f^2 \tilde \phi '' -
\alpha \theta \phi f^2 \tilde R'' + 2\alpha f'\dot \phi \nonumber\\
 4T_{vv} \equiv T_{tt}  + f^2 T_{rr}  + 2fT_{tr}
 &=&\dot \phi ^2  + f^2 \phi '^2  + 2\alpha f^2 \tilde \phi '' - \alpha
\theta \phi f^2 \tilde R'' + 2\dot \phi \left( {f\phi ' - \alpha
f'} \right)
\nonumber\\
   &=&\left( {\dot \phi  + f\phi '} \right)^2  + 2\alpha f^2 \tilde \phi ''
- \alpha \theta \phi f^2 \tilde R'' - 2\alpha f'\dot \phi
\label{vac2}
\end{eqnarray}

 In order to study different vacuum states in the metric
 (\ref{nostra}),  we need to fix integration constants  in
 (\ref{fi}).  Let us first  consider the case where there is no energy
flux at any point.  In our case this condition leads to

 \begin{equation}
  T_{tr}\left(\, r\,\right)  = 0 \Rightarrow \dot
  \phi  = 0 \longrightarrow A=0
 \end{equation}

 With this choice of constant $A$, light-cone components are equal and given by

 \begin{equation}
 4T_{uu} \equiv 4T_{vv}
 =\left( {f\phi '} \right)^2  + 2\alpha f^2 \tilde \phi '' - \alpha \theta
\phi f^2 \tilde R''
 \end{equation}

 In order to be able to calculate explicitly the above components, let us
rewrite the scalar field $\tilde\phi$ as

 \begin{eqnarray}
 \tilde \phi  &=& \phi ^0  + \left( {e^{\theta\square }  - 1} \right)\phi ^0
 = \phi ^0  + \sum\limits_{n = 1} {\frac{{\left( {\theta\square }
\right)}}{{n!}}} ^n \phi ^0\nonumber\\  &=& \phi ^0  - \alpha
\theta \sum\limits_{n = 0} {\frac{{\left( {\theta\square }
\right)}}{{(n + 1)!}}}
^n R\nonumber \\
 \tilde \phi ' &=& \phi ^{0\,\prime }  - \alpha \theta \sum\limits_{n = 0}
{\frac{{\theta ^n }}{{(n + 1)!}}} (\Re ^n )' \\
  \tilde \phi '' &=& \phi ^{0\,\prime \prime }  - \alpha \theta
\sum\limits_{n = 0} {\frac{{\theta ^n }}{{(n + 1)!}}} (\Re ^n
)''\nonumber
\\
 \Re ^n  &=& \square^n R \label{nc}
  \end{eqnarray}
 where we have used equation of motion $\square\phi^0=-\alpha R$.
 Now, we can calculate the field derivatives at the horizon. Keeping in mind
that $f(r_H)\equiv f_H=0$, the only components  surviving in
(\ref{vac2}), when  $r=r_H$, are

 \begin{eqnarray}
 \left(\, f^2 \tilde \phi ''\,\right)_{r=r_H}  &=&\left(\,  f^2 \phi^{0\,
 \prime\prime} \,\right)_{r=r_H}  =  - \left(\, \alpha f_H^{\prime\, 2}  +
 B f_H^\prime\,\right) \nonumber\\
 \left(\,  f^2 \tilde \phi ^{\prime\,
 2}\,\right)_{r=r_H}  &=& \left(\, \alpha f'_H  + B \,\right)^2
\end{eqnarray}

The constant $B$   is determined by the condition that there is no
outgoing flux  from  the (future) horizon. This   gives the
Hartle-Hawking vacuum

\begin{equation}
 T_{uu}\vert_{r=r_H}  = B^2  - \left(\, \alpha f'_H \,\right)^2  = 0
\Rightarrow B_{HH} = \frac{\alpha  }{2M_H }\left(\, 1 - 2M_H
^\prime \, \right)
 \end{equation}

   Boulware vacuum is determined by the requirement
  that there is no outgoing flux at  infinity which leads to

  \begin{eqnarray}
 && T_{tr}=0\Rightarrow A =0\\
 && \lim_{r\to \infty}T_{uu}  = \lim_{r\to
 \infty}T_{vv} =  0 \Rightarrow B=0
   \end{eqnarray}

Finally, the Unruh vacuum  describing  an evaporating black hole,
originated by a collapsing body,   requires no incoming flux from
past infinity, and no flux escaping out of the future horizon.
These conditions lead to

\begin{eqnarray}
&& T_{vv}|_{r\rightarrow\infty}  = 0 \Rightarrow \left(\dot \phi
+ f\phi
\right)|_{r\rightarrow\infty}=0\rightarrow A  =  - B \nonumber\\
 && T_{uu}|_{r=r_{H}}  = 0 \Rightarrow B = \frac{{\alpha f'}}{2}
 =\frac{\alpha }{{4M_H }}\left(\, 1 - 2M_H ^\prime \, \right)
\end{eqnarray}

\begin{table}
\caption{We summarize the values of the integration constants
leading to three different vacua. }
\begin{tabular}{llccc}
Vacuum  & Constant  $A$ & Constant $B$ & Physical meaning \\
\colrule
&    &    & \\
Boulware  & $A=0$ & $B=0$ & Vacuum Polarization\\
Hartle-Hawking & $A=0$ & $B=  \frac{\alpha }{2M_H }(1 - 2M_H
^\prime  )$ & Black Hole in a Thermal
Bath\\
Unruh & $A=-B$ & $B=\frac{\alpha }{4M_H }\left(\, 1 -
2M_H ^\prime \, \right) $ & Evaporating Black Hole\\
&    &    & \\
\colrule
\end{tabular}
\label{table1}
\end{table}

The values of the $A$, $B$ integraion constants, and the
physical meaning of the corresponding vacua, are summarized
in (\ref{table1}).
\subsection{Short-distance behavior }
It is interesting to investigate the behavior of the
energy-momentum tensor at short distances $r^2/4\theta<<1$ where
non-commutative effects should be important.  In this regime
 the energy-momentum tensor components are given only by local part

\begin{eqnarray}
T_{tt}^{loc} &=&\frac{A^{2}+B^{2}}{2}-\frac{1}{2}\alpha
^{2}f^{\prime
2}+2\alpha ^{2}ff^{\prime\prime }. \\
T_{rr}^{loc} &=&\frac{A^{2}+B^{2}}{2f^{2}}-\frac{1}{2f^{2}}\alpha
^{2}f^{\prime 2} \\
T_{rt}^{loc} &=&\frac{AB}{f}
\end{eqnarray}

Using the expansion of incomplete $\gamma$ function given in the
Appendix B one finds $f\simeq 1-\Lambda r^{2}$,
  where $\Lambda =\frac{M}{3\sqrt{\pi }\theta ^{3/2}}$. Thus, at short
distances the metrics is the De Sitter one and
 the components of energy-momentum tensor are

\begin{eqnarray}
T_{tt}^{short} &\simeq &\frac{A^{2}+B^{2}}{2}-4\alpha ^{2}\Lambda
+O\left( r^2
\right) \\
T_{rr}^{short} &\simeq &\frac{A^{2}+B^{2}}{2}
 +O\left( r^2
 \right) \\
T_{rt}^{short} &\simeq &AB
 +O\left( r^2
 \right)
\end{eqnarray}

The result shows that coordinate uncertainties remove the
curvature singularity at the origin leading to a regular behavior
in, otherwise, divergent quantities in all three vacua.

\subsection{Long-distance behavior }
Long distance behavior  shows no significant difference form
ordinary results.   This follows from the fact that
non-commutative effects are exponentially small and the metric
approaches ordinary Schwarzschild solution at distance larger than
$\sqrt\theta$.   Indeed, one finds

\begin{equation}
f_{  r^2/4\theta  >>1}\left(\, r\,\right) \approx
1-\frac{2M}{r}+\frac{2M}{\sqrt{\pi\theta}}e^{-r^2/4\theta}\equiv
f_S\left(\, r\,\right)+\frac{2M}{\sqrt{\pi\theta}}e^{-r^2/4\theta}
\label{flarge}
 \end{equation}

Energy momentum tensor  $T_{\mu\nu}^{loc.}$  is  calculated with
$f$ given by (\ref{flarge}) .  One obtains usual Schwarzschild
form of the vacuum expectation value of $T_{\mu\nu}$  \cite{balbi}
plus exponentially damped corrections. This is in agreement with
the general belief that the non-commutative effects modify, in a
sensible way,  only  short-distance physics. At large distances
spacetime looks as a smooth manifold and non-commutative effects
are   experimentally non-observable.

\section{Conclusions}

We have shown how to modify  the heat kernel  asymptotic expansion
to include the quantum coordinate fluctuations leading to an
intrinsic de-localization of spacetime events at short distances.
As a result of the unavoidable   position  uncertainty, no UV
divergencies are present. This result led us to propose an
alternative way to compute the trace anomaly by using an IR
regulator.  The resulting trace anomaly acquires a non-local
character . \\
Trace anomaly has been already used in $2D$ as an essential
ingredient to determine the vacuum expectation value of the energy
momentum tensor.   Following these ideas, we have calculated
corrections to the energy momentum tensor mean value in the
Boulware, Hartle-Hawking and Unruh vacua. To carry out this
calculation we have used  $rt$ section of a recently found,
noncommutative inspired, Schwarzschild-like solution of the
Einstein equations.  The underlaying motivation is that quantum
coordinate fluctuations influence both matter \textit{and }
geometry itself.  The metric found in \cite{last} incorporates the
latter effect.  It turns out that the short distance behavior of
the vacuum expectation values of $T_{\mu\nu}$ is now  regular .
The metric itself is of deSitter form near the origin, thus,
resolving the curvature singularity of the Schwarzschild solution.

\section{Appendix A}
In this appendix we prove the relation (\ref{dirac}).
\begin{equation}
e^{\theta\,\Box_x}\equiv \sum_0^\infty \frac{1}{n!}\, \left(\,
\theta \, \Box_x \,\right)^n
\end{equation}

\begin{eqnarray}
e^{\theta\, \Box}\, e^{i k x} &&= \sum_0^\infty \frac{1}{n!}\,
\left(\, \theta \, \Box_x \,\right)^n\, e^{i k x}\nonumber\\
&&= \sum_0^\infty \frac{1}{n!}\,
\left(\, -\theta \, k^2 \,\right)^n\, e^{i k x}\nonumber\\
&&=e^{-\theta\, k^2}\, e^{i k x}
\end{eqnarray}

\begin{eqnarray}
e^{\theta\, \Box_x}\,\delta^{(d)}\left(\, x-y\,\right) &&=
e^{\theta\, \Box}\,\int \frac{d^dk}{\left(\, 2\pi\,\right)^d}\,
e^{i\, k\, \left(\, x-y\,\right)}\nonumber\\
&&= \int \frac{d^dk}{\left(\, 2\pi\,\right)^d}\, e^{-\theta\, k^2}
e^{i\, k\, \left(\, x-y\,\right)}\nonumber\\
&&=\frac{1}{\left(\, \pi\,\right)^d}\,\left(\,
\frac{2\pi}{\theta}\,\right)^{d/
2}\,  e^{-\left(\, x-y\,\right)^2/4\theta\, }\nonumber\\
&&\equiv \rho_\theta\left(\, x\ ,y\,\right)
\end{eqnarray}

In the same way, one can show that the same widening effect occurs
when $e^{\theta\, \Box_x}$ operator is applied to a finite width
Gaussian:

\begin{equation}
e^{\theta\, \Box_x} \frac{1}{\left(\, 2\pi\, s\,\right)^{d/2} }\,
\exp\left( -\frac{x^2}{4s}\, \right)= \frac{1}{\left[\,
2\pi\,\left(\, s +\theta  \,\right) \,\right]^{d/2} }\, \exp\left(
-\frac{x^2}{4\left(\, s +\theta \,\right) }\, \right)
\end{equation}

\section{Appendix B.\,\,\bf Properties of incomplete Gamma functions}

Definitions of incomplete lower $\gamma$ and upper $\Gamma$  functions\\

\begin{eqnarray}
\gamma\left( \frac{n}{2}+1\ , x^2\right) &\equiv&
\int_0^{x^2} dt\, t^{n/2} e^{-t}\nonumber\\
\Gamma\left( \frac{n}{2}+1,x^2\right) &\equiv& \int_{x^2}^{\infty}
dt \,t^{n/2} e^{-t}=\Gamma\left(\, \frac{n}{2}
+1\,\right)-\gamma\left(\, \frac{n}{2}+1\ , x^2\,\right)
\end{eqnarray}
Integral and differential properties of incomplete $\gamma$ function\\

\begin{eqnarray}
 &&\int_0^r  \frac{dx}{x^{n + 1}  }\gamma \left(\, \frac{n}{2}
 +1\ , x^2 \,\right)  =  -
\frac{1}{2}\frac{\gamma \left(\, \frac{n}{2} \ , r^2\,\right)}{r^n
}
\nonumber\\
 &&\int {d^n } x\rho _\theta  (x^2 ) =
 M\frac{\gamma \left(\,\frac{n}{2}\ ,R^2\,\right)}{\Gamma
 \left(\, \frac{n}{2}\,\right)} \nonumber\\
 && \gamma^{\prime}\left(\, \frac{n}{2},
 x^2\, \right)=2\, x^{n-1}\, e^{-x^2}\nonumber\\
 && \gamma\left( \frac{n}{2}+1,\frac{r^2}{4\theta}\,\right)=
 \frac{n}{2}\gamma\left( \frac{n}{2},
\frac{r^{2}}{4\theta }\right)-\left(\, \frac{r}{2\sqrt{\theta}}\,
\right)^n\, e^{-r^{2}/4\theta }
\end{eqnarray}
 Long and short distance behavior of incomplete $\gamma$ functions\\
 \begin{eqnarray}
 && \gamma \left(\, \frac{n}{2} \ , x^2 \,\right)\vert_{x>>1}
 =\frac{2}{n}x^n e^{ - x^2} \left[ \, 1 - \frac{2}{{n + 2}}x^2  + \frac{2}{{n +
2}}\frac{2}{{n + 4}}x^4  + \dots \right] \nonumber\\
&&  \gamma \left(\, \frac{3}{2}\ ,\frac{r^2}{4\theta}
  \,\right)\vert_{\frac{r^2}{4\theta}<<1}
\approx  \frac{r^3}{12\sqrt{\theta^3}} \left(1 -
\frac{7}{20}\frac{r^2}{\theta}\right)\nonumber\\
&& \Gamma\left(\, \frac{n}{2}\ , x^2\, \right)\vert_{x>>1} =
x^{n-2}\
e^{-x^2}\left[1+\left(\frac{n}{2}-1\right)\frac{1}{x^2}+\left(\frac{n}{2}-1\,
\right)
 \left(\frac{n}{2}-2\right)\frac{1}{x^4}+\dots\right]\nonumber\\
 &&\gamma\left( \, \frac{3}{2}\ ,
\frac{r^2}{4\theta}\, \right)\vert_{\frac{r^2}{4\theta}>>1}
 =\frac{\sqrt{\pi}}{2}-\Gamma\left( \frac{3}{2}
\frac{r^2}{4\theta}\right)\vert_{\frac{r^2}{4\theta}>>
1}\approx\frac{\sqrt{\pi}}{2}
 +\frac{1}{2}\frac{r}{\sqrt{\theta}}\,e^{-\frac{r^2}{4\theta}}
\end{eqnarray}

Long and short distance behavior of the metric

\begin{eqnarray}
f&=&\left( 1-\frac{2M\left( r\right) }{r}\right)\nonumber\\
&=&1-\frac{2M}{r}-\frac{4m}{\sqrt{\pi}\,r}\Gamma\left( \frac{3}{2}
\frac{r^2}{4\theta}\right)\nonumber\\
f^{\prime }&=&\frac{2M(r)}{r^{2}}-\frac{2M^{\prime }(r)}{r}\nonumber\\
&=&\frac{4M}{\sqrt{\pi}r^2}\left[\gamma\left( \frac{3}{2}
\frac{r^{2}}{4\theta
}\right)-\frac{r^3}{4\theta^{3/2}}e^{-r^{2}/4\theta
}\right]\nonumber\\
f^{\prime \prime }&=&-\frac{4M(r)}{r^{3}}+
\frac{4M^{\prime }(r)}{r^{2}}-\frac{2M^{\prime \prime }(r)}{r}\nonumber\\
&=&-\frac{8M}{\sqrt{\pi}r^3}\left[\gamma\left( \frac{3}{2},
\frac{r^{2}}{4\theta
}\right)-\frac{r^5}{16\theta^{5/2}}e^{-r^{2}/4\theta
}\right]\nonumber\\
M\left(r\right) &=&\frac{2M}{\sqrt{\pi }}\gamma\left( \frac{3}{2},
\frac{r^{2}}{4\theta }\right) \nonumber\\
M^{\prime }(r)&=&4\pi r^2\rho_\theta(r)=\frac{M r^{2}}{2\sqrt{\pi
}
\theta ^{3/2}}e^{-r^{2}/4\theta }\nonumber\\
M^{\prime\prime }(r)&=&\frac{M r}{\sqrt{\pi }\theta
^{3/2}}e^{-r^{2}/4\theta
}\left(1-\frac{r^2}{4\theta}\right)\nonumber\\
\end{eqnarray}

\begin{eqnarray}
 f_{ r^2/4\theta >>1} && \approx 1-\frac{2M}{r}+\frac{2M}{\sqrt{\pi\theta}}e^{-
r^2/4\theta}=
f_S+\frac{2M}{\sqrt{\pi\theta}}e^{-r^2/4\theta}\nonumber\\
f_{ r^2/4\theta<<1} && \approx 1- \frac{M}{3\sqrt{\pi }\theta
^{3/2}}r^{2}
 \end{eqnarray}

\end{document}